\documentclass[aip,jcp,reprint,twocolumn,showpacs,amsmath,amssymb,groupedaddress,showkeys,superscriptaddress]{revtex4-1}

\usepackage{bm} 

\usepackage{graphicx,color,float}
\usepackage{subfigure}
\usepackage{dcolumn}
\usepackage{bm,xfrac}
\usepackage{mhchem}
\usepackage{times,mathptmx}
\usepackage{lineno}

\begin{document}

\title{Structure and dynamics of hydrodynamically interacting finite-size Brownian particles in a spherical cavity: spheres and cylinders}

\author{Jiyuan Li}
\thanks{These two authors contribute equally}
\affiliation{Pritzker School of Molecular Engineering, University of Chicago, Chicago, Illinois 60637, USA}
\author{Xikai Jiang}%
\thanks{These two authors contribute equally}
\affiliation{Pritzker School of Molecular Engineering, University of Chicago, Chicago, Illinois 60637, USA}
\affiliation{State Key Laboratory of Nonlinear Mechanics, Institute of Mechanics, Chinese Academy of Sciences, Beijing 100190, China}
\author{Abhinendra Singh}
\affiliation{Pritzker School of Molecular Engineering, University of Chicago, Chicago, Illinois 60637, USA}
\author{Olle G. Heinonen}
\affiliation{Materials Science Division, Argonne .pdfNational Laboratory, Lemont, Illinois 60439, USA}
\affiliation{Northwestern-Argonne Institute for Science and Engineering, Evanston, Illinois 60208, USA}
\author{Juan P. Hern\'andez-Ortiz}
\email{jphernandezo@unal.edu.co}
\affiliation{Departmento de Materiales y Nanotecnolog\'ia, Universidad Nacional de Colombia, Sede Medellin, Colombia}
\affiliation{Colombia/Wisconsin One-Health Consortium, Universidad Nacional de Colombia, Sede Medellin, Colombia}
\affiliation{Pritzker School of Molecular Engineering, University of Chicago, Chicago, Illinois 60637, USA}
\author{Juan J. de Pablo}
\email{depablo@uchicago.edu}
\thanks{To whom correspondence should be addressed.}
\affiliation{Pritzker School of Molecular Engineering, University of Chicago, Chicago, Illinois 60637, USA}
\affiliation{Materials Science Division, Argonne National Laboratory, Lemont, Illinois 60439, USA}

\date{\today}

\begin{abstract}
 The structure and dynamics of confined suspensions of particles of arbitrary shape is of interest in multiple disciplines, from biology to engineering. Theoretical studies are often limited by the complexity of long-range particle-particle and particle-wall forces, including many-body fluctuating hydrodynamic interactions. Here, we report a computational study on the diffusion of spherical and cylindrical particles confined in a spherical cavity. We rely on an Immersed-Boundary General geometry Ewald-like method to capture lubrication and long-range hydrodynamics, and include appropriate non-slip conditions at the confining walls.  A Chebyshev polynomial approximation is used to satisfy the fluctuation-dissipation theorem for the Brownian suspension.
We explore how lubrication, long-range hydrodynamics, particle volume fraction and shape affect the equilibrium structure 
and the diffusion of the particles. It is found that once the particle volume fraction is greater than 10\%, the particles start to form 
layered aggregates that greatly influence particle dynamics.  Hydrodynamic interactions strongly influence the particle diffusion 
by inducing spatially dependent short-time diffusion coefficients, stronger wall effects on the particle diffusion towards the walls, and 
a sub-diffusive regime --caused by crowding-- in the long-time particle mobility.  
The level of asymmetry of the cylindrical particles considered here is enough to induce an orientational order in the layered structure, decreasing the diffusion rate and facilitating a transition to the crowded mobility regime at low particle concentrations.  Our results offer 
fundamental insights into the diffusion and distribution of globular and fibrillar proteins inside cells. 
\end{abstract}

\keywords{Hydrodynamics, Crowded Brownian dynamics, Confinement, Immersed Boundary}
\maketitle

\section{\label{sec:introduction}Introduction}
Diffusion under confinement is central to multiple physical, chemical and biological systems, including colloidal and protein suspensions, devices for particle separation, and transport through membranes. A model system to study the diffusion and structure of highly concentrated particles under confinement could offer insights into the dynamics of crowded macromolecules, such as proteins, inside cells where they typically occupy 20 to 40\% 
of the cytoplasm volume~\cite{Ellis:2001bu}.

Previous studies have shown that crowding between macromolecules affects reaction rates of equilibrium reactions~\cite{Ellis:2001gy,Zhou:2008ht,Anonymous:2008bo,Ellis:2006hs,Minton:2005dm}, and hinders the diffusion of intra-cellular particles~\cite{Dix:2008gy}.
\textit{In vivo} experiments, using fluorescence recovery after photo-bleaching (FRAP) techniques, have reported that {the apparent diffusion coefficient for green fluorescent proteins (GFP) in \textit{E. Coli}'s cytoplasm is about 11 times lower than that in water~\cite{swaminathan1997photobleaching, terry1995molecular, Konopka:2006kf}.}  
Even though a variety of intracellular activities, namely metabolism, cellular homeostasis, signaling, transcription, translation and locomotion, strongly depend on diffusion of the intracellular macromolecules, the mechanisms behind the hindered diffusion are not fully understood. 
A review by Skolnick, discussed several factors that could lead to hindrance, including the viscosity of cytoplasm, steric effects, hydrodynamic interactions (HI), and other short-range interactions between particles~\cite{Skolnick:2016jd}.
A significant effort that relied on Brownian simulations of 50 {types of} macromolecules modeled as spheres inside \textit{E. Coli}'s cytoplasm was able to reproduce the translational diffusion coefficient of GFP~\cite{and:2006cf,McGuffee:2010bi}. The authors demonstrated that steric repulsions cannot explain the hindered diffusion and suggested that electrostatic and other short-range interactions are essential variables to consider. However, short-range interactions modeled as Van der Waals potentials can always be tuned to match experimental results, and these authors were unable to provide conclusive remarks regarding the effects of short- and long-range HI.
In other work, BD simulations considering fluctuating HI have been performed in the bulk~\cite{Ando:2010hg}, confirming that HI plays a nontrivial role in the hindering of macromolecular diffusion. Unfortunately, the mobility variations induced by confinement was not considered in this work.
Confinement,  using a network of wall particles that are constrained by a predefined potential, was included in a subsequent study to represent a cell membrane~\cite{Chow:2015ja}. In that work, 
the non-slip boundary condition was not strictly satisfied, allowing the flow to penetrate the membrane, and driving a tangential component of the particle average velocity.
{
From the hydrodynamic interactions point of view, previous efforts were centered on the study of the dynamics of single or many particles immersed in an unconfined viscous fluid~\cite{jeffery1912form, jeffery1915steady, oseen1927neuere, stimson1926motion,happel1960motion,o1969exact,sherief2017general,lamb1924hydrodynamics,happel2012low,kim2013microhydrodynamics,batchelor1970slender} and of particles moving near a wall or confined in a slit and cylindrical geometries~\cite{brenner1986,dean1963slow,o1964slow,o1967slow,goldman1967slow,pozrikidis1994motion,ganatos1980strong1,ganatos1980strong2,swan2011hydrodynamics}.
Recently, a theoretical study was performed by Zia et. al. to model the behavior of a concentrated colloidal dispersion confined in a spherical cavity~\cite{zia2018}. 
Their framework relies on a set of hydrodynamic tensors that capture far- and near-field (lubrication) hydrodynamics between particles and walls. The authors studied the structure and diffusion of hydrodynamically interacting spherical particles confined in a spherical cavity following a Stokesian dynamics (SD) approach~\cite{zia:2016}. 
They found that the confinement, crowding and HI collectively lead to an anisotropic micro-structure, which then induce position-dependent and anisotropic short- and long-time dynamics.} That study was, however, limited to spherical particles.

We have developed an efficient computational framework to perform BD simulations of arbitrarily shaped  particles confined in any type of geometry.  We use 
an Immersed-Boundary (IB) method to represent the suspended particles, a 
parallel Finite Element General geometry Ewald-like method (pFE-GgEm)~\cite{Zhao:2017} to calculate the confined Green's functions, 
and a Chebyshev polynomial approximation to satisfy the fluctuation-dissipation theorem. In this work, we use this methodology to study 
how steric repulsion, short- and long-range hydrodynamic interactions, confinement, particle volume fraction and particle shape affect the structure and the diffusion of 
spherical and cylindrical finite-size particles confined in a spherical cavity.  The cylinders are selected to break the three-dimensional
symmetry of the particles, a feature that is common in protein structures. 

The paper starts with a description of the particle model, the geometry, the methods and the 
dimensionless variables to characterize the system. The BD and the IB-pFE-GgEm methods are briefly summarized. 
We then proceed to present and discuss the results, including the spatial and orientational ordering of the particles and the short- and long-time diffusion behavior. 
We finish the paper with a summary of the most important findings.

\section{\label{sec:methods}Mathematical model and system } 
Let's consider ${N}$ mono-disperse and semi-rigid particles embedded in a viscous fluid of viscosity $\eta$, confined in an spherical cavity of radius $R$. Under a zero Reynolds number condition, the ${N}$-body force/torque balance on the 
particles is:
\begin{equation} \label{eqn-solids}
\mathbf{F}^H + \mathbf{F}^B + \mathbf{F}^{C} + \mathbf{F}^{EV} + \mathbf{F}^{ext} = 0,
\end{equation}
where $\mathbf{F}^{H}$ is the $6{N}$ hydrodynamic force/torque vector, $\mathbf{F}^{B}$ is the 
Brownian force/torque vector, $\mathbf{F}^{C}$ is the force/torque vector containing configurational forces,
$\mathbf{F}^{EV}$ represents the force/torque vector due to excluded volume interactions, and 
$\mathbf{F}^{ext}$ includes all external forces/torques. 

The evolution of the
suspended particles, from Eqn.~\eqref{eqn-solids}, is carried out through the grand mobility or
resistance tensors that relate the hydrodynamic forces/torques to the translational and 
rotational velocities of the particles~\cite{pozrikidis,ladyz,power_book}. SD~\cite{Brady1988,SierouBrady01,brady2010} 
and boundary integral methods (BIM)~\cite{pozrikidis,hernandez_book} have been used extensively to 
solve this ``mobility problem". 
The regularized Stokeslets~\cite{cortez01}, the accelerated BIM~\cite{Kumar:2012ev}, and the Immersed 
Boundary (IB)~\cite{Peskin2002,Atzberger:2007p3828} approaches 
all provide computational efficiency and simplicity, typically to improve (or avoid) the calculation 
of the single- and double-layer hydrodynamic potentials of the suspended particles.  In particular, 
the IB method represents the surfaces of the suspended solids as a distribution of discrete force densities that, together  
with a surface force description and Stokes equations, generate the temporal evolution of the suspended particles. This is the 
approach that we use in this work. 

The surface of each suspended particle is discretized into a set of {$N_\text{IB}$} nodes that constitute a mesh, similarly to
boundary element methods~\cite{hernandez_book}. On the surface nodes, we define structural spring potentials that maintain particle shape, volume and surface. The force balance on each of the $N$ particles is then 
translated into the ${N_\text{IB}}$ surface nodes as follows, 
\begin{equation} \label{eqn1}
\mathbf{f}_\nu^H + 
\mathbf{f}_\nu^B + \mathbf{f}_\nu^{C} + \mathbf{f}_\nu^{EV} = \mathbf{0}, 
\end{equation}
for every node $\nu=1, ..., N_\text{IB}$; where $\mathbf{f}_\nu^H$ is the hydrodynamic force, $\mathbf{f}_\nu^B$ is the Brownian force, $\mathbf{f}_\nu^C$ is the constitutive force and $\mathbf{f}_\nu^{EV}$ is the force from all the excluded volume interactions:  particle-particle and particle--wall. 

Assuming that the probability density for the nodal positions is a continuous density for the 
Fokker-Planck equation~\cite{risken-book}, an equivalent stochastic differential equation for the motion of the nodes 
is written as follows~\cite{Ottinger1996},
\begin{equation} \label{eq:spde_1}
d\mathbf{R} = \left[ \mathbf{M}\cdot\mathbf{F}+
			            \frac{\partial}{\partial \mathbf{R}} \cdot \mathbf{D}
			            \right] dt
			            + \sqrt{2} \mathbf{B} \cdot \mathbf{dW},
\end{equation}
where $\mathbf{R}= (\mathbf{x}_{1}, \mathbf{x}_{2}, \ldots, \mathbf{x}_{N\times N_\text{IB}})$ denotes a {$3N\times N_\text{IB}$} vector containing the spatial coordinates 
of the nodes, $\mathbf{D}=k_BT\mathbf{M}$ is the $3N\times 3N$ diffusion tensor, $k_B$ is Boltzmann constant, $T$ is the 
absolute temperature and $\mathbf{M}$ is the {($3N\times N_\text{IB}$)$\times$ ($3N\times N_\text{IB}$)} mobility tensor. In addition, 
$\mathbf{U}=\mathbf{M} \cdot \mathbf{F}$ contains the $3N$ fluctuating velocities from the hydrodynamic interactions
and $\mathbf{F}$ is the {$3N\times N_\text{IB}$} vector that contains the non-HI and non-Brownian forces on the nodes. 
The divergence of the diffusion tensor $ \frac{\partial}{\partial \mathbf{R}} \cdot \mathbf{D} $ is the drift resulting from the configuration-dependent mobility of the confined particles and $\mathbf{dW}$ is a random vector, the components of which are obtained from a real-valued Gaussian distribution with zero mean and variance $dt$;  $\mathbf{dW}$ is coupled to the diffusion tensor through the fluctuation-dissipation theorem: $ \mathbf{D}=\mathbf{B} \cdot \mathbf{B}^{T} $. 

In IB methods, the force distributions at moving solids are discretized as distributions of regularized point-forces. The 
 ``smoothing" function for the the delta function scales as the distance between the surface nodes that are used to represent 
 the moving particles. Consequently, the structural forces on the particles, $\mathbf{f}_\nu^C$, define a force density as follows
\begin{equation} \label{eq:eqn3}
\boldsymbol{\rho}_\text{IB}^{C}(\mathbf{x})=\sum_{\nu=1}^{N_\text{IB}}
\mathbf{f}^\text{C}_{\nu}(\mathbf{x}_\nu) \delta_\text{IB}(a,\mathbf{x}-\mathbf{x}_{\nu}),
\end{equation}
where $\delta_\text{IB}(a,\mathbf{x})$ is a smoothing function that is a  modified Gaussian (details available in~\cite{Pranay2010}).
{The regularization parameter $\xi_\text{IB}$ in $\delta_\text{IB}$ is related to the characteristic length $h$ for the node spacing on the particle surface, i.e, $\xi_\text{IB}\sim h^{-1} \sim a^{-1}$}. 
The rational behind $\delta_\text{IB}(a,\mathbf{x})$ is to ensure that the regularized force on each node is spread over the length 
scale of the associated surface elements, thereby preventing fluid from penetrating the particle surface. 

\begin{figure} 
\centering
\includegraphics[width=0.9\columnwidth]{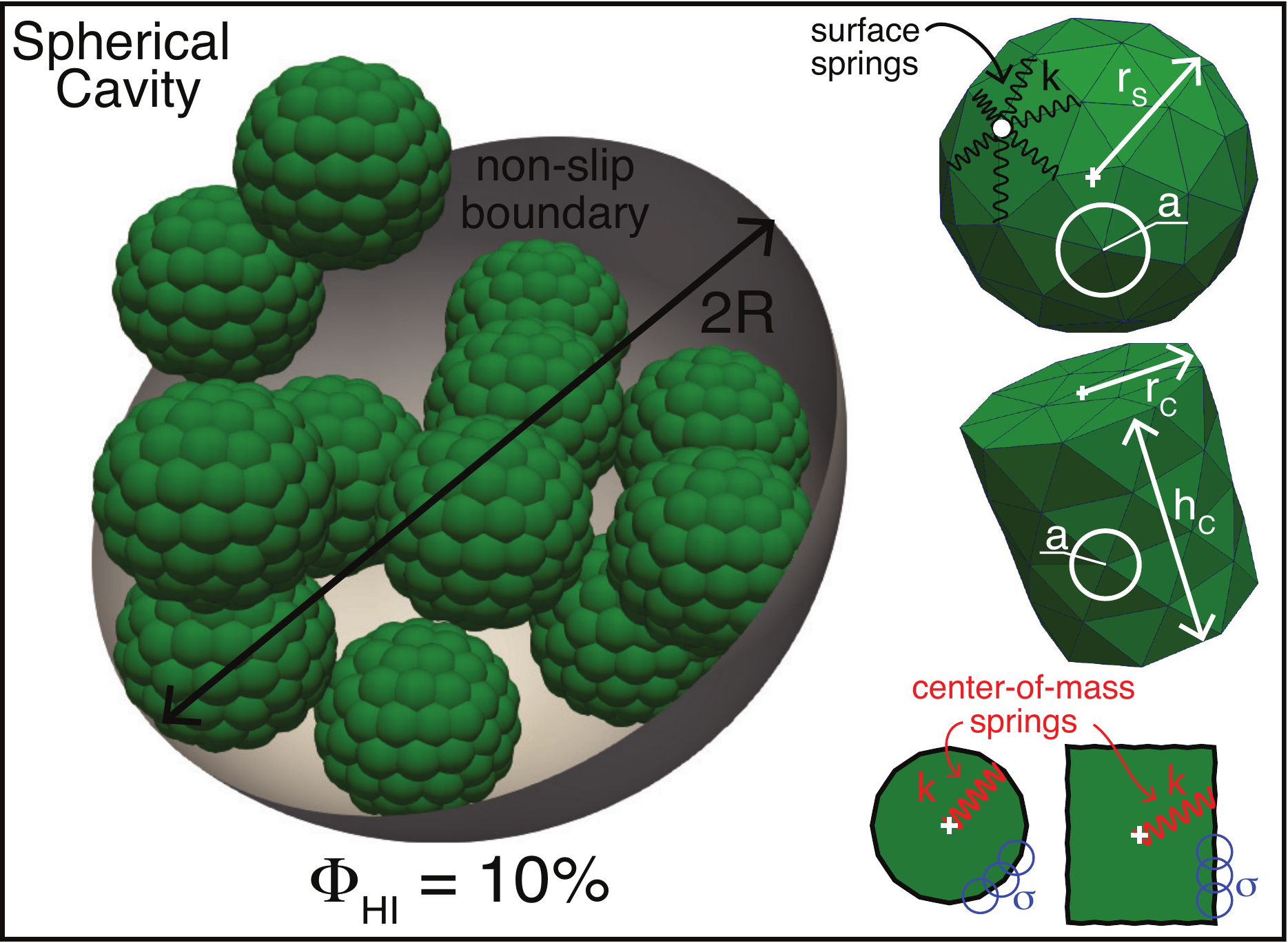}
\caption{Snapshot of the spherical cavity of radius $R$ containing spherical particles with $\Phi_\text{HI} = 10\%$. 
The spherical particles radius is $r_S$, while the size of the cylindrical particles is determined by $r_C$ and $h_C$. 
The surface of the particles is given by a collection of discrete nodes that are connected to six neighbors (``trimesh"), similar
to boundary element discretizations, and with a characteristic node separation of {$a\sim h\sim \xi_\text{IB}^{-1}$}. 
The neighboring nodes are connected with surface springs (black); each node is also connected to the particle 
center-of-mass (red spring). A repulsive Lennard-Jones excluded volume is included on each surface node, shown 
schematically in the particles' cross section by the blue circles. The characteristic size of the repulsion is 
given by $\sigma = 2.2a$.}
\label{fig:system_representation}
\end{figure}

In this work, we consider the particles as ``semi-rigid", where each node on the particle is
linked to its neighboring nodes by elastic springs with a prescribed large stiffness constant. 
The nodes are also connected to the particle center-of-mass by an elastic spring to conserve the desired shape. For simplicity, we assumed
that each link is a linear spring, where the force acting on the point $\nu$ by the point $\mu$ is given by
\begin{equation} \label{eq:eq_linear_spring}
\mathbf{f}^\text{C}_{\nu\mu} = k\left( r_{\nu\mu} - q_{0} \right) \frac{\mathbf{x}_{\nu\mu}}{r_{\nu\mu}}.
\end{equation}
Here $k$ is the spring elastic constant, $q_0$ is the equilibrium spring size for each
specific situation, $\mathbf{x}_{\nu\mu}=\mathbf{x}_\nu - \mathbf{x}_\mu$ and $r_{\nu\mu}=|\mathbf{x}_{\nu\mu}|$. 
For each particle, a spring network is formed (see Fig.~\ref{fig:system_representation}), resulting in an internal nodal force
that resists the deformation of the particles. 
{In addition, at each surface node of each particle a purely repulsive Lennard-Jones (LJ) force is added to account for particle-particle and particle-wall excluded volume interactions.  
This LJ force comes from two contributions: 1) interactions with nodes from a different particle; 2) interactions with the wall.
And it is equivalent to the negative gradient of the LJ potential which is defined by}
\begin{equation} \label{eq:lj}
V^\text{LJ}(r) =  4 k_BT  \left[ \left( \frac{\sigma}{r}\right)^{12}-\left(\frac{\sigma}{r} \right)^6 \right] +k_BT, 
\end{equation}
for $r \le 2^{1/6}\sigma$ and zero otherwise. Here, $r$ is the Euclidean distance between nodes of two different particles 
or between the node and walls, and {$\sigma = 2.2a$ is chosen empirically to guarantee that each surface node has an excluded volume of radius $a$}. 
{The translational and rotational velocity of particles are calculated by integrating the velocity over the surface mesh of each particle, thereby satisfying the force and torque balance.} 
For completeness, we include a validation of our IB approach in the Appendix: 
(i) fulfillment of Stokes' law of particles under confinement, (ii) validation of the fluctuation-dissipation theorem (diffusion of particles), (iii) consistency of the particle shape as a function of the spring constants.

In what follows, we use dimensionless variables for all results and discussions. 
We use $a$ as the characteristic length scale and the nodal spacing diffusion time, $a^2\zeta/k_BT$, as the characteristic time scale. 
We set $k_BT$ as the scale for energy and $k_BT/a$ for the force. The friction coefficient $\zeta$ is related to the fluid viscosity $\eta$ 
and $a$ through Stokes' law: $\zeta=6\pi\eta a$, and the nodal diffusion coefficient is $D_0 = k_BT/\zeta$. 

Our simulation method, denoted by \textbf{IB-pFE-GgEm}, is an $O(N)$ algorithm that {includes} hydrodynamic interactions for confined, large-scale suspensions of finite-size particles of arbitrary shape. Details can be found in our previous work~\cite{Zhao:2017}. Briefly, the $O(N)$ algorithm consists of three major components: (a) pFE-GgEm routines to calculate the Green's function (Stokeslet) for any geometry, (b) Fixman's mid-point algorithm for time integration and (c) the Chebyshev polynomial approximation for the fluctuation dissipation theorem.  The parallel finite element GgEm (pFE-GgEm) routines are built using open source libraries, thereby facilitating usage of our software. The routines can be downloaded at \textcolor{blue}{http://miccomcodes.org} as part of the Continuum-Particle Simulation Suite (COPSS) from the Mid-west Integrated Center for Computational Materials (MICCoM).

\section{\label{sec:results}Results}
\begin{figure} 
\centering
\includegraphics[width=0.9\columnwidth]{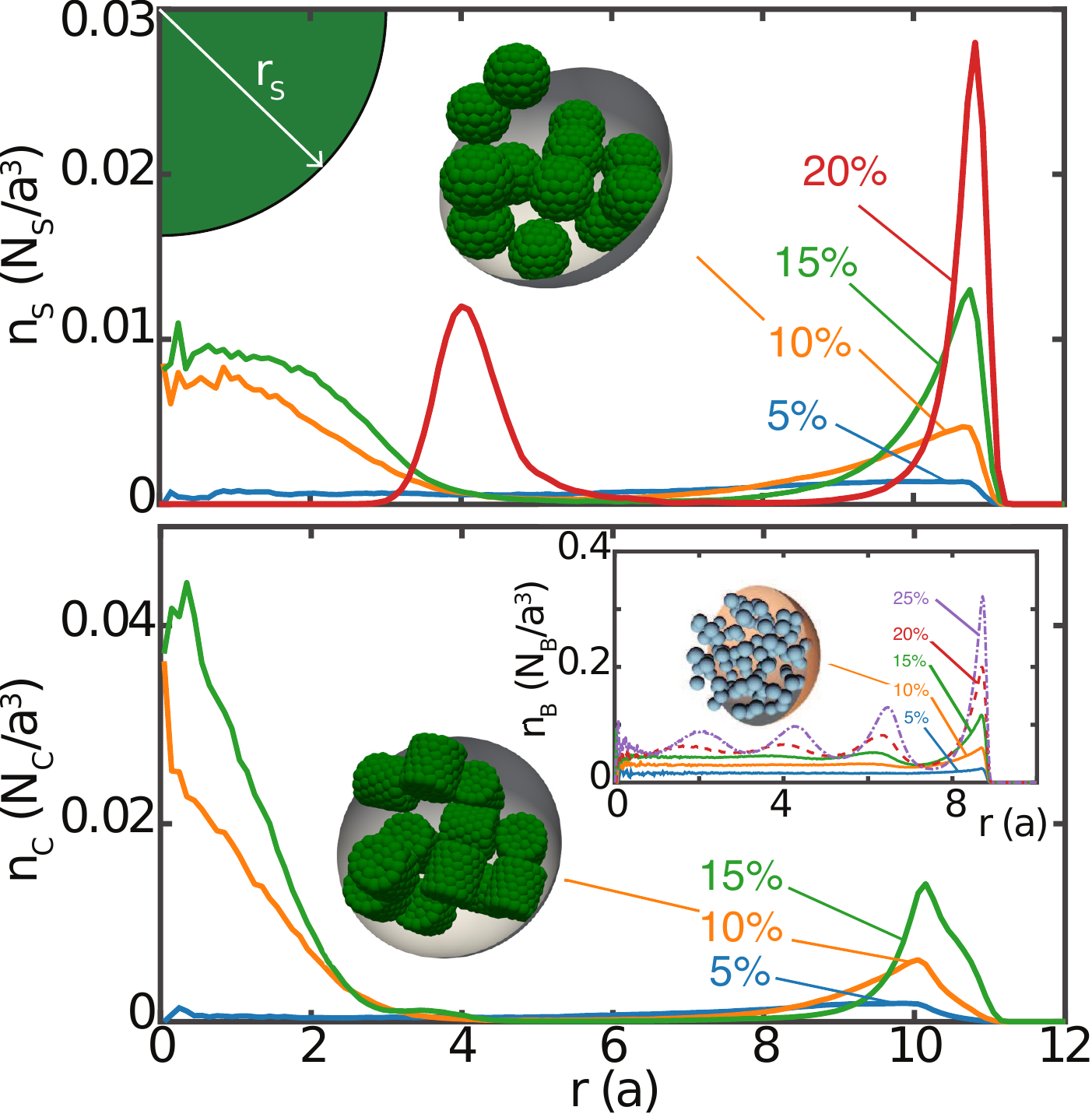}
\caption{Number density of the particles within a spherical cavity of radius $R=15$ as a function of the particle concentration.
(top) spheres with $r_S=3$; and (bottom) cylinders with $r_C=2.62$ and $h_C = 2r_C$. 
Snapshots for $\Phi_\text{HI} = 10\%$ are shown for both systems, while the number density of HI ``beads" is included in the inset. }
\label{fig:particle_density}
\end{figure}

We consider spheres and cylinders, of equal volume, that are suspended in a spherical cavity of radius $R=15$. The particles' radius 
is $r_S = 3$ (volume $V_\text{HI}=4/3\pi r_S^3$), while the cylinders' size is determined by $r_C=2.62$ and $h_C = 2r_C$ 
(volume $V_\text{HI}=\pi r_C^2h_C$). 
Figure~\ref{fig:system_representation} shows the details of the system. 
According to our semi-rigid particle model, there are two ways to define the particle concentration in a cavity of volume $V$. One uses the 
hydrodynamic volume fraction, $\Phi_\text{HI} = {N} V_\text{HI}/V$, and a second one is based on the excluded 
volume effective size, $\Phi_\text{EV} = {N} V_\text{EV}/V$.
{Each surface node has an excluded volume of radius $a$, thus each spherical particle has an excluded volume of
$V_\text{EV} = 4/3\pi (r_S+a)^3$ and each cylindrical particle has an excluded volume of $V_\text{EV}=\pi (r_C+a)^2(h_C+2a)$.} 
We use the hydrodynamic volume fraction as the relevant scale for particle concentration. In this work, 
$\Phi_\text{HI} = [5\%, 10\%, 15\%, 20\%]$ which is equivalent to $\Phi_\text{EV} = [12\%, 24\%, 36\%, 48\%]$. 
{The lower volume concentration selected in the work is $5\%$,  which is the limiting volume fraction between dilute and finite concentration regimes. We found that all the results for dilute systems, starting from infinite dilution, are close to those for the limiting volume fraction of $5\%$}.  
We start our analysis by exploring the structure of the particles using the 
particle number density as a function of radial position within the cavity.  The number density (the probability that 
a particle is at a specific location) is calculated by discretizing the 
 spherical cavity into $m$ bins (spherical shells) with an even spacing in the radial direction.
 The shell radius of the $i\text{-th}$ bin is $b_i = (i+0.5)R/m$. The particle number density $n(r_i)=<N(r_i)/V_i>$, where $N(r_i)$ is the number 
 of particles at shell that is located a distance $r_i$ from the center of the cavity, $V_i$ is the volume of the shell and $<>$ represents an ensemble average over time. Figure~\ref{fig:particle_density} shows the number density of the particles within the cavity as a function of particle concentration. 
Figure~\ref{fig:particle_density}(top) is for spherical particles, while Fig.~\ref{fig:particle_density}(bottom) is for the cylindrical ones. At low concentrations, $\Phi_\text{HI} = 5\%$, the
number density is uniform throughout the cavity and goes to zero once the particles are in contact with the wall.
{Notice that the maximum density is at $r\sim 10.5$, which is smaller than wall contact. In our model, particles can never ``touch" the wall because of the strong repulsive Lennard-Jones potential.}
As the concentration increases, 
the probability of finding particles near the wall increases, {forming a layered structure.}
{At moderate concentrations, $\Phi_\text{HI}=10\%,15\%$, particles start to form the first layer next to the wall, and inner particles prefer to stay near the center of the cavity where the steric effects with the particles in the first layer are the weakest. 
At higher concentrations, $\Phi_\text{HI}=20\%$, particles form a second layer, since there is not enough space at the center to accommodate them. 
Thus the layered 
structure becomes more pronounced, and the layer separation is determined by the particle size.}  
{In the figure, we include the number density of a ``bead" system where particles are represented as spheres with an excluded volume with radius $a$. The spherical cavity has a radius of $R=10a$. In this system, particles interact only through far-field hydrodynamics and steric repulsions. However the layered structure is also observed, becoming a characteristic of highly concentrated confined systems.} 
 
\begin{figure} 
\centering
\includegraphics[width=0.9\columnwidth]{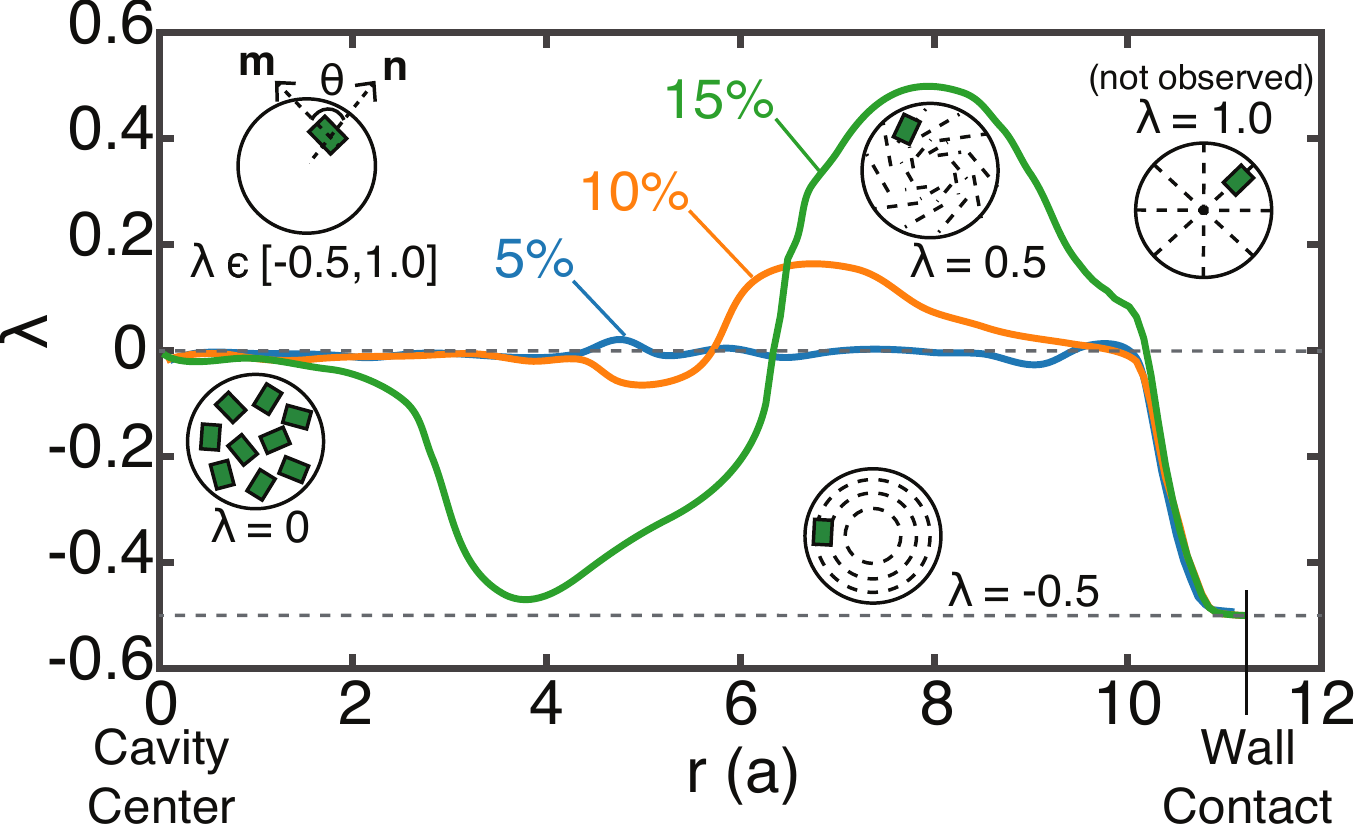}
\caption{Orientational order parameter $\lambda$ of cylindrical particles within a spherical cavity of $R=15$. 
The radius of cylinders is $r_C=2.62$ and the height $h_C=2r_C=5.24$.}
\label{fig:average_lambda}
\end{figure}

Cylindrical particles exhibits an orientational distribution within the cavity. Similar to liquid 
crystalline systems~\cite{tsvetkov1942,martinez2017}, we define an orientational order parameter $\lambda=\frac{1}{2}\langle3\cos ^2 \theta-1\rangle$,
where $\cos\theta=\sfrac{\mathbf{m}\cdot\mathbf{n}}{||\mathbf{m}||\cdot||
\mathbf{n}||}$, $\mathbf{m}$ is the vector parallel to the cylinder's centerline and $\mathbf{n}$ is the vector pointing from the cavity center to the cylinder's center-of-mass. A random/disordered structure is characterized by $\lambda=0$, whereas for ordered morphologies 
 $\lambda=1$, when all cylinders are aligned parallel to the radial direction of the spherical cavity (radial phase), and $\lambda = -1/2$ when all cylinders are aligned perpendicular to the radial direction of the spherical cavity (concentric phase). 
 Figure~\ref{fig:average_lambda} shows the orientational order parameter of cylindrical particles within the cavity as a function of particle concentration. A major result is that for all concentrations the cylinders are in a disordered state at the center of the cavity and
 oriented concentrically near the wall.  For $\Phi_\text{HI}=5\%$, the disordered morphology spans all locations in the cavity, whereas 
 for $\Phi_\text{HI}=15\%$ two concentric regions are observed, separated by a layer where the cylinders are oriented following
  a  $<\theta> =35^o$ or 145$^o$. 
 Interestingly, the radial--centripetal ordering, $\lambda=1$, is never observed; we believe this is due to small aspect ratio of the cylindrical particles. 

We now proceed to analyze the short- and long-scale diffusive behavior in an attempt to delineate the 
consequences of long- and short-range HI on the dynamics of the particles.  The short-time diffusion coefficient and
the mean squared displacement (MSD) of the particles are used to quantify these effects. The particles are suspended in a viscous fluid under zero Reynolds number conditions; they interact with other particles and 
walls through the HI. Recall that under these conditions, momentum transport is infinitely fast~\cite{hernandez-migration,hernandez-nlogn,hernandez-spp,Kounovsky-Shafer2013}.  
{Previous studies of confined suspensions have shown that there are multiple factors that originate from HI that should affect the diffusion (mobility)
 of the particles: (i) the reduction of
the particle mobility due to confinement -- particles in the bulk diffuse faster than in confined geometries --, (ii) the space dependent mobility due to the non-slip 
conditions at the walls -- particle diffusion is zero at the walls --, and (iii) the decrease of particle mobility as the concentration increases -- there is 
an interplay between lubrication and long-range HI that becomes important as the average inter-particle distance decreases~\cite{brenner1961slow,o1967slow,swan2011hydrodynamics,zia2018}. We seek to quantify and determine the consequences of such factors in a cavity enclosure using our model.}

\begin{figure} 
\centering
\includegraphics[width=0.9\columnwidth]{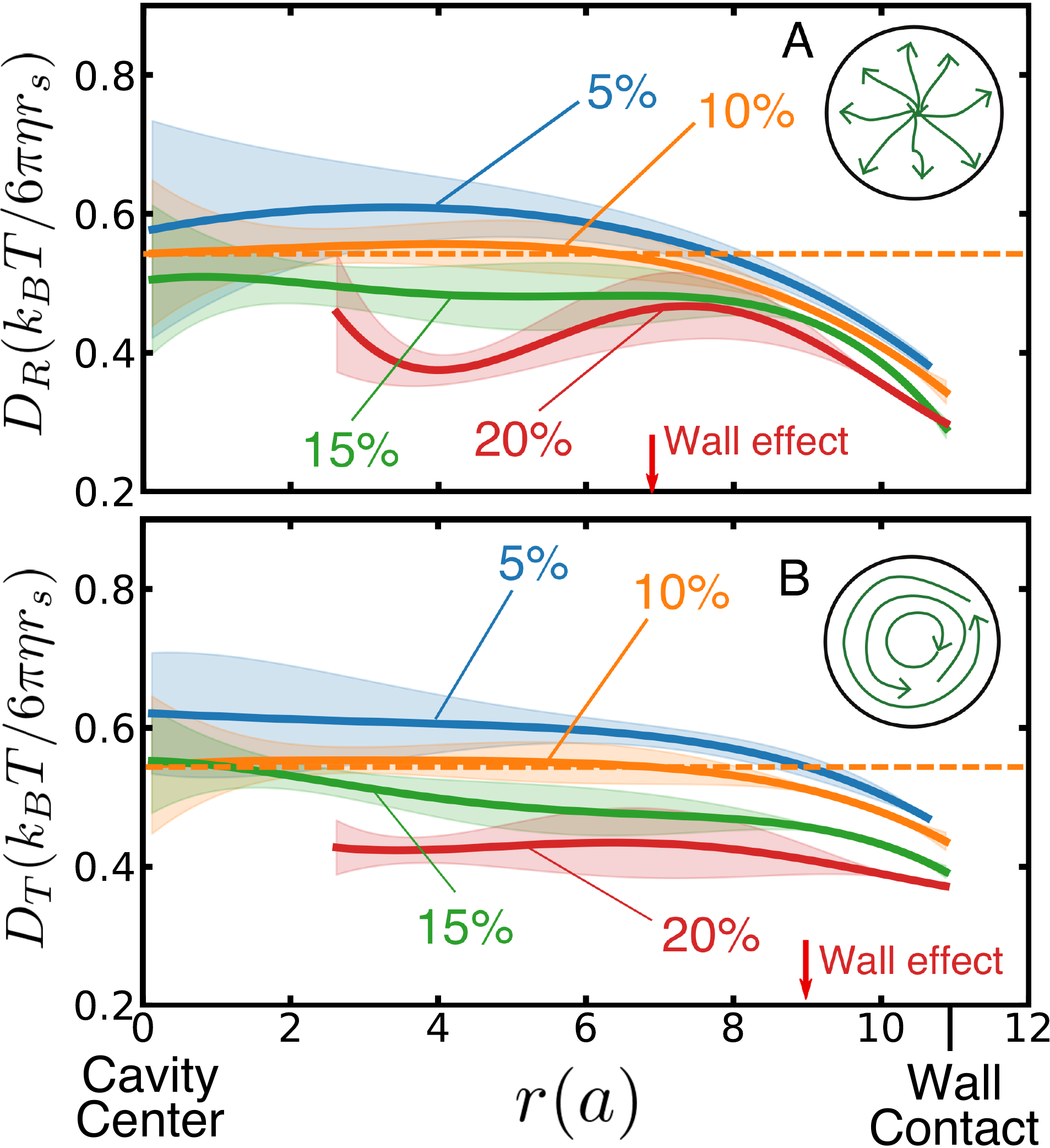}
\caption{Short-time diffusion coefficients for sphere particles ($r_S=3$) that are confined in a 
spherical cavity with $R=15$: (top) Radial diffusivity and (bottom) Tangential diffusivity. The coefficients are normalized by the diffusivity of spherical  
particles in bulk at infinite dilution, $k_BT/(6\pi\eta r_S)$. The orange dashed line represents the averaged
``inner" diffusivities for $\Phi_\text{HI}=10\%$. The filled shadow area around each curve represents their respective statistical error. 
{The diffusion coefficients for $r/a<2.3$ at $\Phi_\text{HI}=20\%$ are missing because there are not enough particles appearing in this zone for the diffusivity measurement due to the layered structure.}} 
\label{fig:particle_diffusion_shortime}
\end{figure}

We start by measuring the particles' radial and tangential short-time diffusivities within the cavity as a function of particle concentration.  
These transport coefficients are calculated from the relation between MSD and time from Stokes--Einstein following a 
directional decomposition~\cite{Lara2011}, i.e. 
\begin{eqnarray}
\langle{\Delta \mathbf{x}_\text{R}^2}\rangle(r_i) &=& 2D_\text{R}(r_i)dt, \\    
\langle{\Delta \mathbf{x}_\text{T}^2}\rangle(r_i) &=& 4D_\text{T}(r_i)dt, 
\end{eqnarray}
for $t \to 0$ and where $\Delta \mathbf{x} = \mathbf{x(t+dt)} - \mathbf{x(t)}$, 
the radial displacement $\Delta \mathbf{x}_\text{R}=\Delta\mathbf{x}\cdot\mathbf{x}/|\mathbf{x}|$, 
the tangential displacement $\Delta \mathbf{x}_\text{T} = \Delta\mathbf{x} - \Delta\mathbf{x}_\text{R}$, 
$D_\text{R}(r_i)$ and $D_\text{T}(r_i)$ are the instantaneous radial and tangential short-time diffusivities at a distance $r_i$ from the center of the cavity, and $dt$ is an infinitesimal time interval. $D_\text{R}(r_i)$ and $D_\text{T}(r_i)$ are blocked averaged during a typical simulation at each shell, and then averaged over independent simulations.  Figure~\ref{fig:particle_diffusion_shortime} summarizes the diffusion coefficients within the cavity 
for spherical particles as a function of concentration. 
{In the figure, the diffusivities are normalized by the bulk value, which is defined as single particle diffusivity at infinite dilution.} 

We find that the confinement hinders the particle diffusion in both directions. The highest value for the particle diffusivity at $\Phi_\text{HI}=5\%$ is around 60\% of the bulk value.  The lower particle mobility is directly related to the long-range character of the HI and the non-slip 
conditions at the walls. As the particle concentration increases, lubrication forces begin to dominate and particle diffusion 
decreases monotonically. In addition, the short-time coefficients are not constant within the cavity, showing sudden decrease as the particles
approach the wall.  For $\Phi_\text{HI}=10\%$, we calculated the averaged ``inner" coefficients, which are represented
by the orange dashed lines in Fig.~\ref{fig:particle_diffusion_shortime}.
{Importantly, the decrease in particle mobility at the walls has a stronger 
effect on the radial diffusion coefficient, indicated by i) the radial particle mobility decreases at $r=7$, when compared with the tangential diffusivity 
that ``feels" the presence of the walls at $r = 9$, ii) the radial diffusivity decreases by 0.2 from $r(a)=7$ to the wall contact, whereas the tangential diffusivity only decreases by about 0.12.}
For a spherical particle near the walls, it is easier to diffuse concentrically than radially. 
Finally, note that for $\Phi_\text{HI}=20\%$ the coefficients adopt a non monotonic character, which is correlated with the layered structure of
particle density in the cavity.
{The diffusivity data for $r(a)<2.3$ is missing for the case of $\phi_\text{HI}=20\%$ because the short-time diffusion coefficient at a given location is measured on particles that appear at that location; in other words, the diffusion coefficients are only measured in regions where there is a finite concentration of particle (see Fig.~\ref{fig:particle_density}(top)).}
{Similar observations have also been reported by Zia et. al. using a Stokesian dynamics (SD) approach~\cite{zia2018}.} 

\begin{figure} 
\centering
\includegraphics[width=0.95\columnwidth]{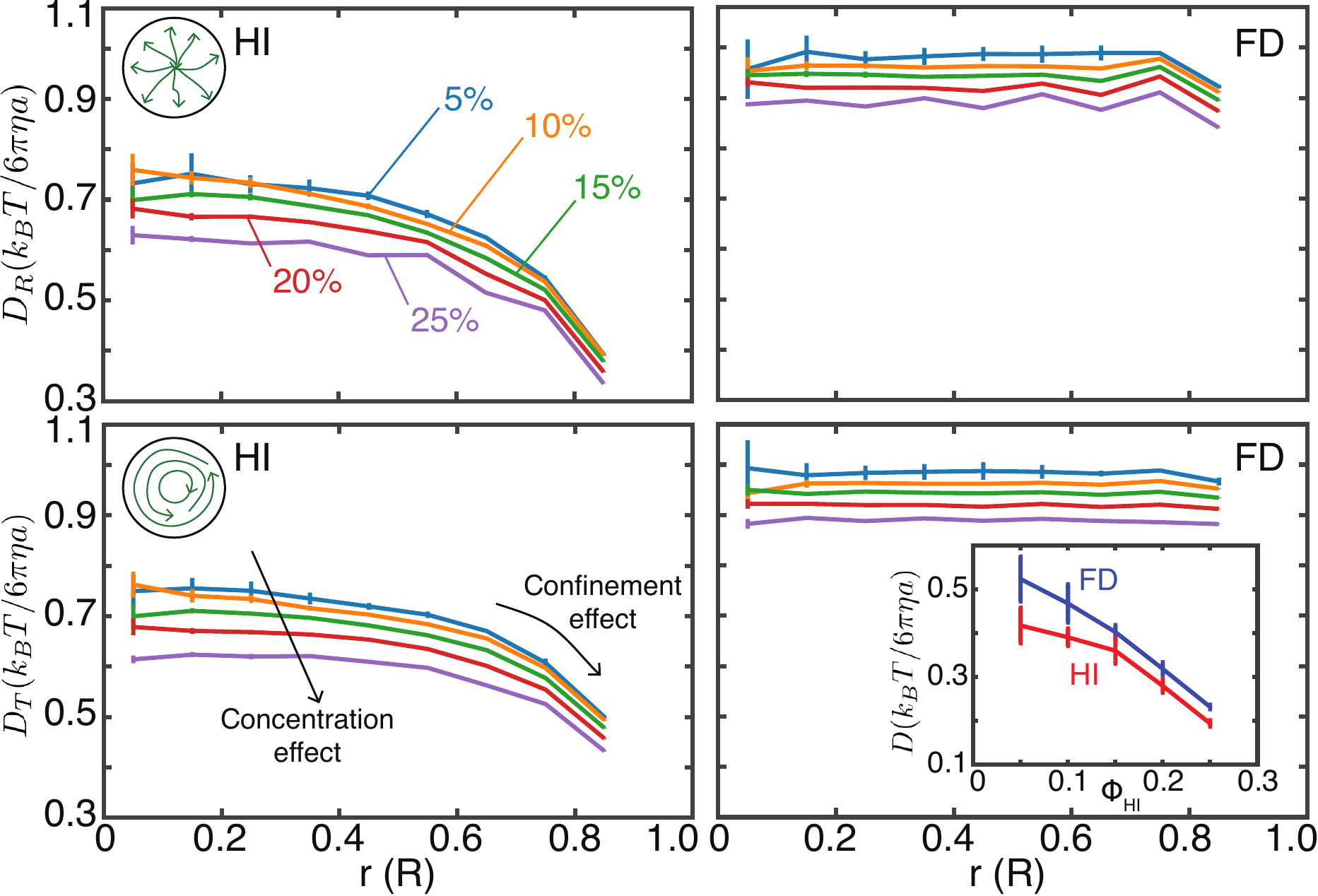}
\caption{Short-time diffusion coefficients for point-particles (``beads", $r_B=1$) that are confined in a 
spherical cavity with $R=10$: Radial and tangential diffusivities for HI (left) and free-draining (right) particles. 
The inset shows the long-time diffusion coefficient as a function of the particle concentration.}
\label{fig:bead_diffusion_shorttime}
\end{figure}

{In Fig.~\ref{fig:bead_diffusion_shorttime} we include the short-time diffusion coefficients of spheres with an excluded volume confined to a spherical cavity, with and without long-range HI (no lubrication forces).} 
Our intention is to change the level of the particle description to isolate the HI contributions.  Free draining point-particles do not undergo a
space dependent diffusion and the concentration perturbs the diffusion rate only weakly. For HI point particles, the mobility decreases 
monotonically as the particles approach the walls and the diffusion rate has a stronger dependence on concentration. Consequently, 
long-range HI and the zero mobility at the walls are responsible for the non-uniform particle diffusion inside the cavity.  Lubrication, on the other hand,
imposes a directionality on the short-time mobility -- radial diffusion is different than concentric diffusion near the walls -- and correlates the particle diffusion with the
layered density at high volume fractions. 
{The observed behavior is consistent with the experimental findings~\cite{carbajal2007asymmetry, eral2010anisotropic}}.

\begin{figure} 
\centering
\includegraphics[width=\columnwidth]{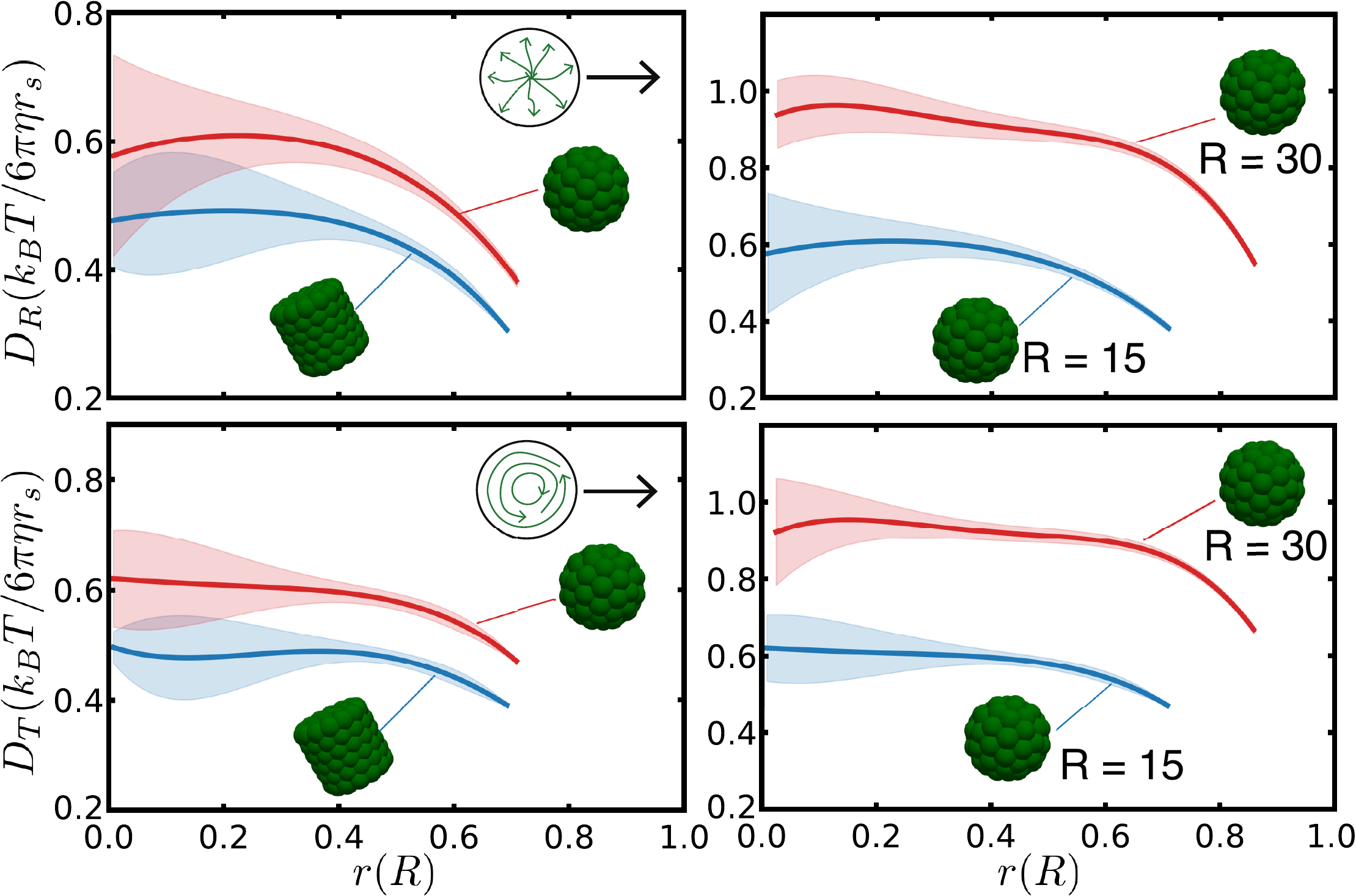}
\caption{Short-time diffusion coefficients for: (left) cylindrical particles with $r_C=2.62$ and $h_C=2r_C=5.24$ confined in a 
spherical cavity with $R=15$ and (right) spherical particles with $r_S=3$ confined in a spherical cavity with $R=30$. 
The particle concentration is $\Phi_\text{HI}=5\%$ and the results for spherical particles with $r_S=3$ confined in 
a spherical cavity with $R=15$ are included for comparative purposes. The filled shadow area on each curve represents their respective statistical error.}
\label{fig:shorttime_diffusion}
\end{figure}

It is of interest to compare the short-time diffusion coefficient between spherical and cylindrical particles and to validate the effects of the level of confinement. Figure~\ref{fig:shorttime_diffusion} includes the short-time diffusion coefficients for
cylindrical particles (left) in a cavity with $R=15$ and for spherical particles (right) in a cavity with $R=30$.  In the figure, $\Phi_\text{HI} = 5\%$ and
the results for spherical particles in a cavity with $R=15$ are included for reference.  
Similar to the spherical particles, the cylinders exhibit position-dependent diffusion coefficients, and radial diffusion is affected strongly by the presence of the walls
when compared to tangential diffusion.  Interestingly, the shape of the cylindrical particles has an important effect on the
rate of diffusion. Recall that the cylinders and spheres have the same volume, and that the diffusion coefficients are normalized by the bulk diffusivity of spherical 
particles. Consequently, the short-time diffusion of particles of equal volume is decreased when the geometrical symmetry is broken.  Finally, decreasing the level of
confinement does not change the qualitative behavior of the short-time mobility but, as the confinement decreases, the inner diffusion coefficients approach the bulk
value. 

Finally, we examine the long time diffusive behavior. We use a generalized Stokes-Einstein relation where the 
MSD is linearly correlated with a mobility coefficient following a power law by:
\begin{equation}
\big\langle \big(  \mathbf{R}(t)-\mathbf{R}(0)  \big)^2 \big\rangle = M t^\alpha,
\end{equation}
where $\mathbf{R}$ is the $3{N}$ particle coordinate vector, $M$ is the particle mobility coefficient and $\alpha$ is the power law exponent
that characterizes the type of particle transport.  {For isotropic diffusion}, if $\alpha = 1$ the particles are diffusive and $M=6D$, where $D$ is the particle long-time diffusion coefficient.  If $\alpha \neq 1$, the transport of Brownian particles is said to be in the anomalous diffusion regime: $\alpha<1$ is sub-diffusive, 
while $\alpha > 1$ is super-diffusive. For confined systems, as $t \to \infty$, the walls impose long-time restrictions on the mobility along the 
confined direction. Therefore, for particles confined in a spherical cavity, the MSD should reach a plateau 
on a time scale corresponding to the particle diffusion time over the cavity size. 

\begin{figure} 
\centering
\includegraphics[width=0.95\columnwidth]{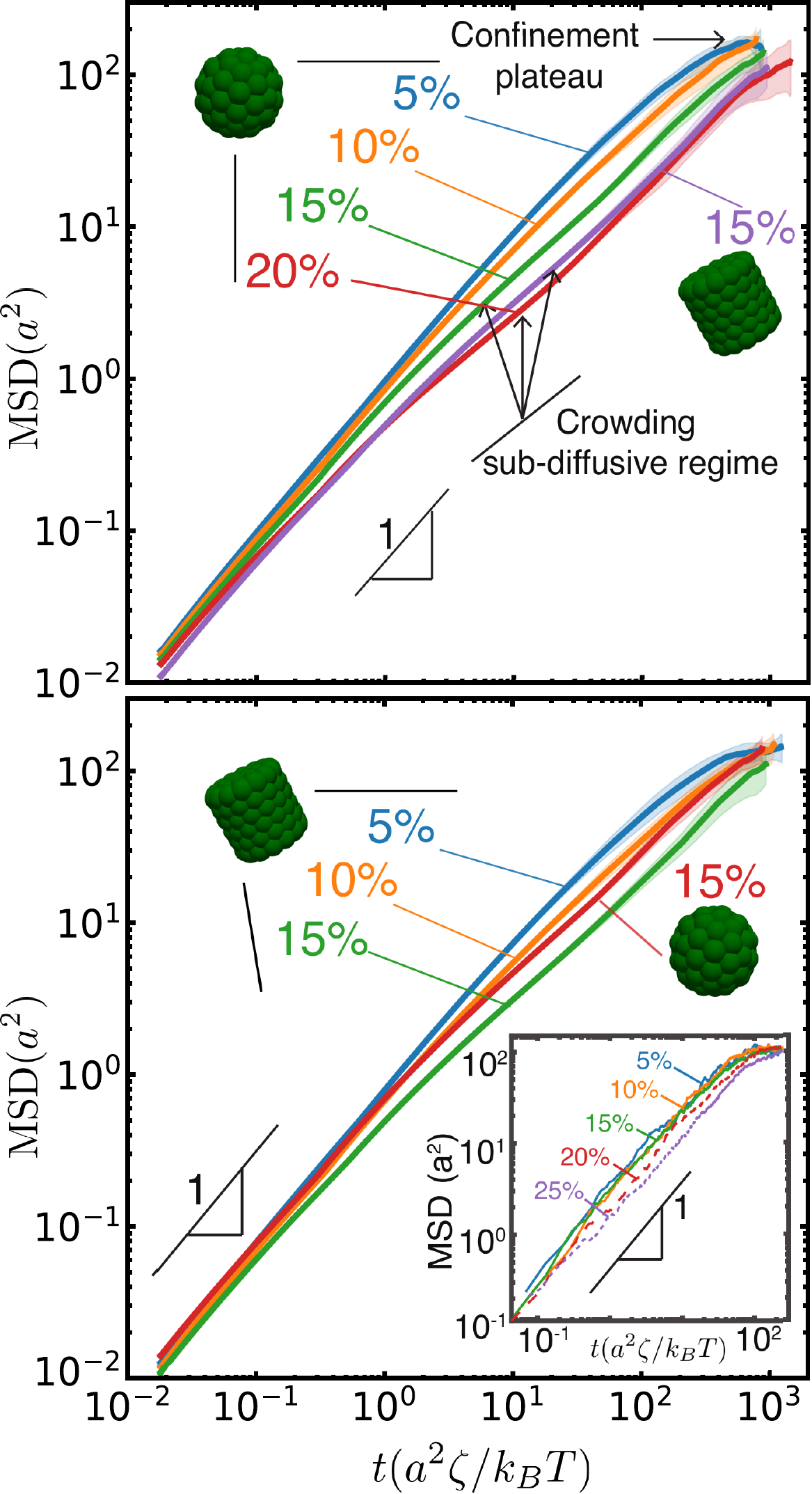}
\caption{Mean square displacement as a function of time for finite-size particles suspended in a spherical particle of size $R=15$. 
(top) spheres with a radius $r_S = 3$ and (bottom)  cylinders with $r_C = 2.62$ and $h_C=2r_C$.
In the inset is the evolution of the mean square displacement for point-particles (``beads") suspended in a spherical 
cavity of radius $R=10$.  In each figure, the MSDs for $\Phi_\text{HI} = 15\%$ of the other particle shape are 
included for comparison purposes.  The filled shadow area around each curve represents their respective statistical error.}
\label{fig:particle_diffusion_longtime}
\end{figure}

Figure~\ref{fig:particle_diffusion_longtime} shows the MSD of spherical (top) and cylindrical (bottom) particles in a cavity 
of $R=15$ as a function of particle concentration.
{The time scale in Fig.~\ref{fig:particle_diffusion_longtime} is in $a$ diffusion time unit, $a^2\zeta/k_BT$, where $\zeta=6\pi\eta a$.}  In the figures, we include the results for $\Phi_\text{HI}=15\%$ of the other
particle shape for comparison purposes.  The MSDs are collected from ten independent simulations with different random seeds; the
particles were able to diffuse more than 300 {$a$} diffusion times. 
{As expected, the MSD exhibits diffusive behavior as $t \to 0$ and a plateau when MSD$\sim 200 \sim R^2$ at $\tau_R$, where $\tau_R$ is characteristic particle diffusion time over the cavity.
At low concentrations, the diffusive behavior, for spheres and cylinders, spans from $t=0$ to $t = \tau_R$ (for $5\%$ spheres, $\tau_R\sim300$ in $a^2\zeta/k_BT$ units).} 
As the particle concentration is increased, the diffusion rate decreases; it is correlated with the short-time behavior, as indicated by the monotonic
shift of the MSDs in Fig.~\ref{fig:particle_diffusion_longtime}.  For spherical particles at $\Phi_\text{HI} \ge 15\%$, there 
is clear sub-diffusive regime over more than two decades that increases as the concentration is increased.  Interestingly, 
the sub-diffusive regime for cylindrical particles starts at $\Phi_\text{HI} = 10\%$.  Note that the MSD for cylindrical particles
at $\Phi_\text{HI} = 15\%$ is almost equal to the MSD for spherical particles at $\Phi_\text{HI} = 20\%$.  {In the inset, we have included the MSD of spheres with excluded volume but no lubrication, to show how the sub-diffusive regime is never observed when only long-range HI is included, thereby suggesting that the sub-diffusive regime is a lubrication effect.}
It is then natural to attribute this anomalous diffusion to crowding, driven by short-range HI. 
This regime is then characterized by the diffusion-to-crowding transition $\tau_\text{DtC}$ time, and by the power law exponent $\alpha$. 
In this regime, there are two major features: $\tau_\text{DtC}$ and $\alpha$ decrease as the particle concentration increases. 
In particular, $\tau_\text{DtC} = 20$ and $\alpha = 0.85$ for cylinders at $\Phi_\text{HI}=10\%$, while 
$\tau_\text{DtC} = 5$ and $\alpha = 0.65$ for cylinders at $\Phi_\text{HI}=15\%$. 
After the highly concentrated systems enter the crowding regime, and given the fact that there are 
three dimensional restrictions in their mobility as $t$ approaches $\tau_R$, the particles transition from the slow rate
sub-diffusive behavior to the plateau. In Appendix ~\ref{app:zia}, we show that our mean square displacements for spheres agree with the isotropic mean square displacement analysis reported in literature~\cite{zia2018, zia:2016}, both qualitatively and quantitatively. {Ref.\onlinecite{zia2018} reported both sub-diffusive and super-diffusive regimes at intermediate time scales, where the latter was only observed for high-concentration systems.}

\section{\label{sec:conclusions}Conclusions}
We have used an Immersed Boundary approach to study the structure and dynamics of suspended spherical and cylinderical particles confined in a spherical cavity. 

At low concentrations, the particle number density distribution is uniform in the interior of the cavity. As the concentration increases, a layered
structure appears. Cylindrical particles exhibit
a random orientation at low concentrations, and at the center of the cavity, for all concentrations.  Excluded volume interactions at the wall 
force the cylinders to orient concentrically. Interestingly, at high concentrations the layered morphology of the cylinders is correlated with 
concentric layers that are separated by a layer of cylinders oriented at an average angle of 35$^o$ with respect to the radial direction. 
Cylinders are never found forming radial morphologies within the spherical cavity. 

We used a multiple hydrodynamical description of the suspended particles to determine the specific influence of
hydrodynamic interactions during the dynamic of the particles (diffusion/mobility): (i) free draining point-particles, (ii) long-range hydrodynamic ``beads" and
(iii) finite size particles considering lubrication and long-range HI.  We found that long-range HI leads to a position-dependent diffusion 
of the particles inside the cavity; the particles diffuse faster near the center of the cavity and slower near the walls. The HI also 
decrease the global mobility of the suspended particles, when compared with their diffusion in the bulk. The increase in particle concentration
also results in a decrease of the particles' diffusion coefficients; this effect is observed for free-draining and HI particles. However,
the concentration decrease of the diffusion rate is stronger when HI are considered.  Lubrication forces, or short-range HI, influence the dynamics of highly concentrated suspensions; they generate a direction dependent diffusion, where
particles diffuse at a lower rate when moving towards the walls than when moving parallel to the walls.  The non-slip conditions at the walls, i.e. zero particle mobility, work synergistically with lubrication forces, resulting in an stronger wall dependence of the diffusion coefficients in the radial direction. 

{
Regarding the long--time dynamics, lubrication gives rise to a sub-diffusive regime
at high particle concentrations. The sub-diffusive regime, characterized by the diffusion-to-crowding transition time and the mobility power law exponent, becomes more prominent as the concentration increases. 
}

Introducing cylindrical particles has two major consequences: (i) cylindrical particles have lower short-time diffusion coefficients and (ii) the crowding 
regime is observed at lower concentrations compared with spheres of equal volume. These observations suggest that the shape of bio-molecules, particles and polymers could determine their mobility and diffusion inside cells and tissues. 

\section*{Acknowledgments}
The development of fast computational codes for simulations of nanoparticles interacting through hydrodynamic and polarization interactions are supported by the Department of Energy, Basic Energy Sciences, Division of Materials Research, through the MICCoM center. The calculations on particle segregation and transport presented in this work are supported by the Department of Energy, Basic Energy Sciences, Division of Materials Research, through the AMEWS center.
\appendix

\section{Suspended Brownian spheres with the IB-pFE-GgEm}
\begin{figure} 
\centering
\includegraphics[width=0.9\columnwidth]{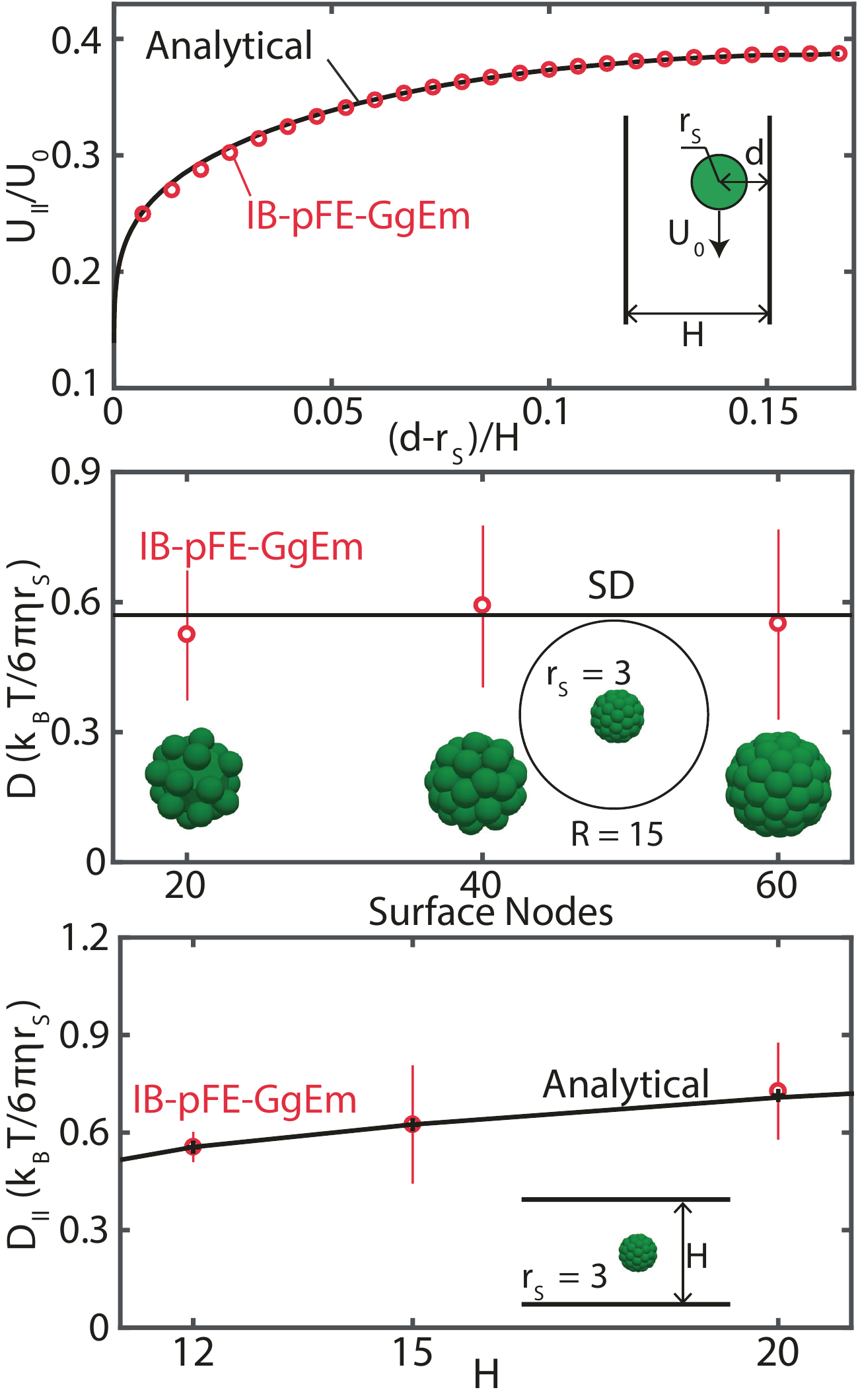}
\caption{Sphere representation with the Immersed Boundary method: 
(top) Normalized sedimenting velocity of a spherical particle between two parallel walls as a function of the distance between 
the particle and the nearest wall. Analytical data are taken from Ref.~\onlinecite{brady2010}. 
(center) Diffusion coefficient of a spherical particle with $r_S=3$ that is confined in a spherical cavity of size $R=15$.
SD data are taken from Ref.~\onlinecite{zia2018}. 
(bottom) $xy-$plane diffusion coefficient for a spherical particle with radius $r_S=3$ in the center of a slit with heights $H=12$, $15$, and $20$.
 Analytical data are taken from Ref.~\onlinecite{de2016finite}.}
\label{fig:sedimentation_validation}
\end{figure}

There are three important validations that are in order to verify the Immersed Boundary approach that we used in this work.
We start by verifying the fulfillment of Stokes' law by measuring the sedimentation velocity of a spherical particle that is
confined between two parallel walls. Analytical values for this velocity are extracted from previous works~\cite{happel2012low,ganatos_weinbaum_pfeffer_1980,ganatos_pfeffer_weinbaum_1980,davis2003,jones2004,bhattacharya2005hydrodynamic,brady2010}. For the specific IB-pFE-GgEm calculation, the particle radius $r_S = 5$ and the wall distance is $H=15$.
Periodic boundary condition (PBC) is enforced in the unconfined directions, which are set to a length of $200$ ($\gg H$) to avoid the influence of
the particle periodic images.  The surface of the spherical particle is discretized using 119 surface nodes, resulting in a nodal separation 
between $h_\text{min}=1.33408$ and $h_\text{max}=1.90539$ and a smoothing parameter $\xi_\text{IB}=1/0.76h_\text{min}$. For 
the calculation we used a GgEm parameter $\alpha_\text{GgEm} = 0.2$ and a mesh resolution with a spacing of $1/\sqrt{2}\alpha$.
The resulting mesh was  $60\times60\times6$ HEX20 elements with 324,886 degrees of freedom. 
The particle is initially located at (0, 0, $d$) between the two parallel walls and it moves parallel to the walls under a sedimenting force 
with a body force density of (0, 0, 1). The particle's sedimenting velocity ($U_{||}$) is calculated by averaging particle's velocities over 100 time steps.
Figure~\ref{fig:sedimentation_validation}(top) shows the sedimenting velocity, normalized by Stokes' velocity $U_{||}/U_0$, as a function of normalized location ($(d-R)/H$). According to the results, the IB-pFE-GgEm provides an excellent agreement with analytical values. 

In addition to Stokes' law, it is important to verify that our combined approach, between the IB-pFE-GgEm, Fixman's mid-point 
algorithm and the Chebyshev polynomial approximation, is satisfying the Fluctuation-Dissipation Theorem. 
A sphere that is confined in a spherical cavity and between parallel walls offers a scenario to validate the connection  
between the diffusion and the fluctuating tensors and the proper calculation of the mobility gradients inside the confined geometries. 
First, we used three different particle discretizations, using 20, 40 and 60 nodes, for the sphere of size $r_S=3$ that is confined in 
a spherical cavity of size $R=15$. Figure~\ref{fig:sedimentation_validation}(center) shows the comparison between 
the short-time diffusion coefficient of the sphere computed through our IB-pFE-GgEm algorithm and SD 
algorithm previously reported in literature~\cite{zia2018}. These results suggest that even with poor surface descriptions, as long as the 
IB parameter is appropriately chosen, the IB-pFE-GgEm follows the correct fluctuating short-time behavior. Finally, 
the long-time diffusion for a sphere confined in a slit as a function of the separation between the walls is 
shown in Fig.~\ref{fig:sedimentation_validation}(bottom).
Analytical and numerical solutions for this coefficient had been well assessed in the literature~\cite{de2016finite,Lin2000PRE,LIN2000Colloid,Dufresne2001}.
For the IB-pFE-GgEm, the particle is initially located at the mid-plane between two parallel walls. During the particle's Brownian movement, its motion is restricted to the plane of symmetry.
The surface of the spherical particle is discretized using 20 surface nodes 
($h_\text{min}=2.19826$, $h_\text{max}=2.52608$ and $\xi_\text{IB}=1/0.35h_\text{min}$). The slit mesh resulted in 
$60\times 60\times 4$ HEX20 elements with 228,872 degrees of freedom.
To calculate the error bars of the MSD, five independent simulations for each confinement ratio are performed with a constant time step of 0.002~\cite{Tarantino231}. 
The diffusion coefficients from the IB-pFE-GgEm shows an excellent agreement with analytical and numerical results.

\section{\label{app:elasticity}Rigidity of the suspended particle}
\begin{figure} 
\centering
\includegraphics[width=0.9\columnwidth]{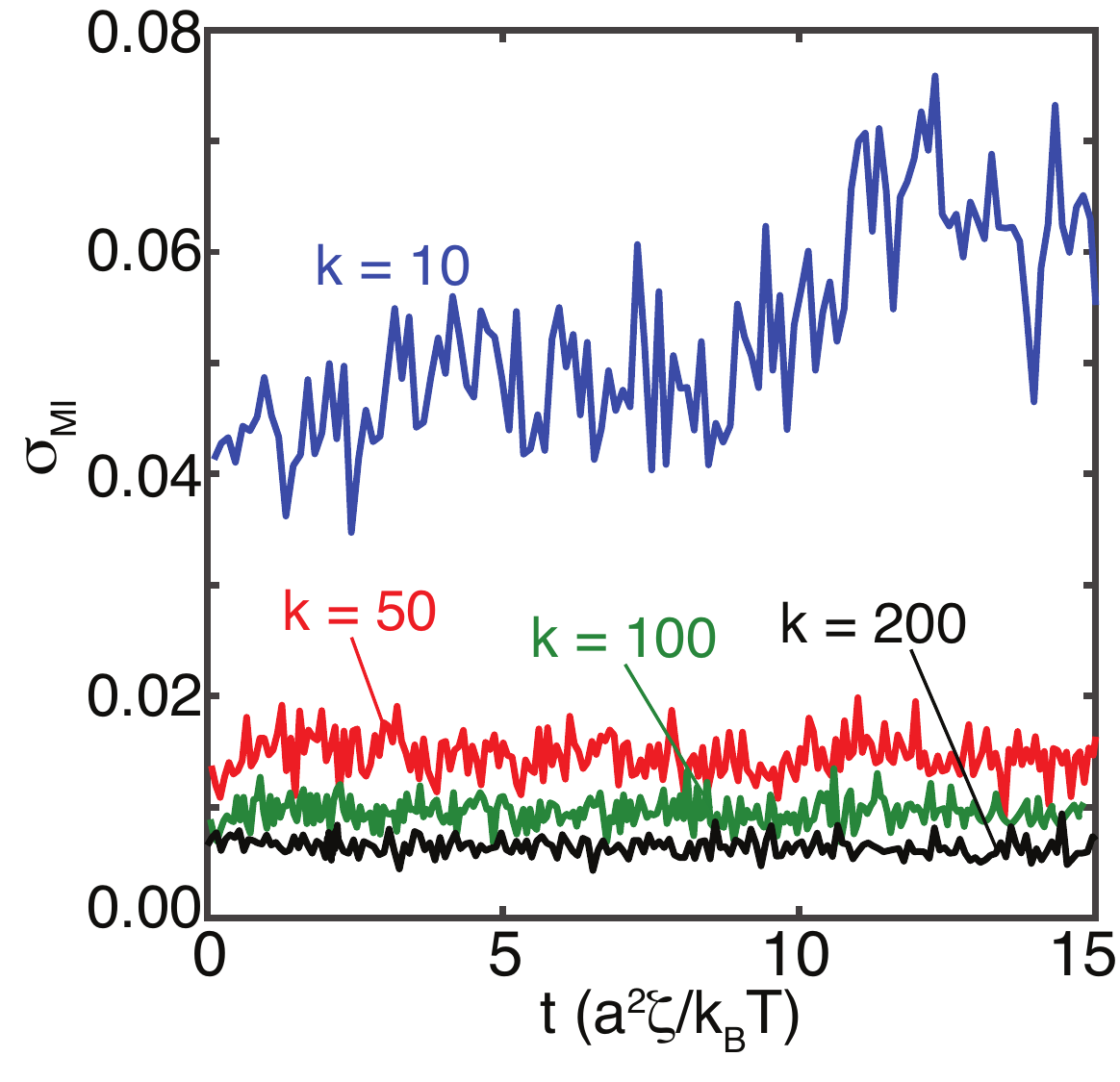}
\caption{Time evolution of the standard deviation of the sphere moment of inertia as a function of the spring stiffness. 
The particle volume fraction is $\Phi_\text{HI}=20\%$, the particle radius is $r_S=3$ and the number of surface nodes on each particle is 20.}
\label{fig:particle_elasticity}
\end{figure}

In our Immersed Boundary description of the finite-size particles, each surface node is linked to its neighboring nodes and the center-of-mass through  linear springs with a prescribed stiffness constant $k$. This parameter controls the stiffness (shape) of particles. If the springs are too weak, particles are deformable and special care must be done to forbid fluid penetration. On the other hand, if the springs are too stiff, the forces acting on surface nodes will be too large, requiring very small time steps to ensure numerical stability.  We performed simulations of 
spherical particles that are confined in a spherical cavity at high concentrations varying the particle stiffness. 
We measured the particles' moment of inertia for each particle $\nu$ as follows:
\begin{equation}
I_\nu(t) = \sum_{j=1}^{N_\text{S}}{r_{j}(t)^2},
\label{eq:moi}
\end{equation}
where $I_\nu(t)$ is the moment of inertia of particle $\nu$ at time $t$ and $r_{j}(t)$ is the distance between node $j$ on surface and 
the particle center-of-mass. Rigid spheres will have an equal and constant moment of inertia. On the other hand, our semi-rigid particles
will show a variation in $I_\nu(t)$.   Figure~\ref{fig:particle_elasticity} shows the averaged standard deviation of moment of inertia as a function
of time for particles with different $k$ for $\Phi_\text{HI}=20\%$.  The standard deviation is calculated by 
\begin{equation}
\sigma_\text{MI}(t) =  \sqrt{ \frac{\sum_{\nu=1}^{{N}}{\big(I_\nu(t)-I_\nu(0)\big)^2}}{{N}}},
\label{eq:moi_mse}
\end{equation}
where $I_\nu(0)$ is the moment of inertia of particle $\nu$ at time $0$, which corresponds to a perfect sphere. 
As one can observe from the figure, in the case that $k=10$, the moment of inertia jumps from 0.05 to 0.06 at $t\approx10$, indicating $k=10$ is not stiff enough to maintain particle shape. As the stiffness is increased the shape variations decrease. In this paper, we used a 
$k=200$ for all simulations.

\section{\label{app:zia}Comparison of MSD with literature results}



The equilibrium structure and diffusion in concentrated hydrodynamically interacting spherical particles confined in a spherical cavity was studied using SD by Aponte-Rivera {\it et al.}\cite{zia2018}. There are differences between methods in our and their work such as particles models, HI models, etc. However, one can draw both quantitative and qualitative agreements between results from two methods.
Here we present the comparison of isotropic MSD at different volume fractions (Fig.15a in Ref.~\onlinecite{zia2018} and the top panel in Fig.\ref{fig:particle_diffusion_longtime} in this work).
Qualitatively, a short-time diffusive regime and a long-time plateau for systems at all concentrations are observed in both works. At intermediate times, a sub-diffusive region emerges for highly concentrated systems only.
 {To make a semi-quantitative comparison, we convert the MSD data in Ref.~\onlinecite{zia2018} to those with the units used in our work at the same level of excluded volume crowding}. 
 Specifically, the length and time reported by Aponte-Rivera {\it et al.} is normalized by the radius of a sphere with radius $r_s$ and the characteristic diffusion time $t_0$ of a sphere with radius $r_s$, respectively ($t_0=r_s^2/D_0$, where $D_0=k_BT/6\pi \eta r_s$). By multiplying the time, MSD, and volume fraction in the their work by $r_s^3$, $r_s^2$, and $(r_s/r_s+a)^3$ ($r_s=3a$), respectively, we can compare the isotropic MSD directly.
For example, for the system with the highest concentration in Ref.~\onlinecite{zia2018} ($40\%$ in their work and $17\%$ after conversion to our unit; the converted value will be reported in round brackets in the following), the short-time diffusive regime lasts from t $\approx 10^{-3}$ ($10^{-2}$) to $\approx10^{-1}$ ($10^0$) while the MSD increases from $10^{-3}$ ($10^{-2}$) to $10^{-1}$ ($10^{0}$); the intermediate sub-diffusive regime lasts from t $\approx 10^{-1}$ ($10^0$) to t $\approx 10^{1}$ ($10^2$) while the MSD increases from $10^{-1}$ ($10^0$) to $10^0$ ($10^1$); the MSD reaches the plateau $10^1$ ($10^2$) at t $\approx 10^2$ ($10^3$). Good agreement is found between our MSD data and those reported in Ref.~\onlinecite{zia2018}.


\bibliography{reference}

\end{document}